\documentstyle[preprint,aps]{revtex}

\def\journal#1, #2, #3, 1#4#5#6{
    {\sl #1~}{\bf #2}, #3 (1#4#5#6)}

\def\prd{\journal Phys. Rev. D, }
\def\prl{\journal Phys. Rev. Lett., }

\def\cmp{\journal Comm. Math. Phys., }
\def\np{\journal Nucl. Phys., }
\def\pl{\journal Phys. Lett., }
\def\annp{\journal Ann. Phys. (N.Y.), }

\def\ijmp{\journal Int. Jour. Mod. Phys., }

\def\etal{{\it et al.}}
\def\cs{Chern-Simons }
\def\beq{\begin{equation}}
\def\eeq{\end{equation}}
\def\LL{{\cal L}}
\def\qu{q_{U(1)}}
\def\qr{q_{ren}}
\def\qs{q_{SU(2)}}
\def\qrs{\qr^{SU(2)}}
\begin{document}
\preprint{UdeM-LPS-TH-93-150, hep-th/9306027}

\title{Spontaneous Symmetry Breaking and the Renormalization
of the Chern-Simons Term}

\author{Avinash Khare,\footnote{permanent address: Institute of Physics,
Sachivalaya Marg, Bhubaneswar, 751005, India.}
R. B. MacKenzie, P. K. Panigrahi\footnote{permanent address:
Department of Physics,  University of Hyderabad, Hyderabad, 500134, India.}
and M. B. Paranjape}
\address{Laboratoire de Physique Nucl\'eaire, Universit\'e de
Montr\'eal C.P. 6128, succ. centreville, Montr\'eal, Qu\'ebec, Canada, H3C 3J7}
\maketitle

\begin{abstract}
\widetext

We calculate the one-loop perturbative correction to the coefficient of the
\cs term in non-abelian gauge theory in the presence of Higgs fields, with
a variety of symmetry-breaking structures. In the case of a residual $U(1)$
symmetry, radiative corrections do not change the coefficient of the \cs
term. In the case of an unbroken non-abelian subgroup, the coefficient of
the relevant \cs term (suitably normalized) attains an integral correction,
as required for consistency of the quantum theory. Interestingly, this
coefficient arises purely from the unbroken non-abelian sector in question;
the orthogonal sector makes no contribution. This implies that the
coefficient of the \cs term is a discontinuous function over the phase
diagram of the theory.

\narrowtext \end{abstract}
\newpage

Yang Mills theories in 2+1 dimensions have attracted much attention in
recent years.  This is primarily because of the possibility of adding a new
topological term to the action, the \cs (CS) term \cite{1,desjactem},
which has had diverse
applications from condensed matter physics to pure mathematics.  The
possibility of particles obeying arbitrary  statistics, anyons \cite{any}, can
be elegantly formulated by using a CS term\cite{csterm}.  Anyons are known to
play an important role in the fractional quantum Hall effect\cite{fqhe}, and
provide a mechanism for superconductivity\cite{anysup}. The limit where the
action is the pure CS term\cite{purecs} results in a topological field
theory\cite{topol}.  The only observables of this theory are Wilson loops,
whose
expectation values give rise to knot invariants.  Considering such a theory
with
a non-compact gauge group, for example the 2+1 dimensional Poincare group,
gives
a consistent quantum theory of 2+1 dimensional gravity\cite{2dg}.
Three-dimensional field theories are the high temperature limits of
corresponding 3+1 dimensional field theories\cite{hit}. Thus parity violating
theories in 3+1 dimensions, like the Standard Model, will in general contain
the
CS term in their effective actions in the high temperature limit.  Therefore it
is interesting to study 2+1 dimensional Yang Mills theory with the
CS term.

For a non-abelian gauge theory with gauge group ${\cal G}$, the CS action,
although invariant under small gauge transformations, changes by an integer
multiple of $8\pi^2\mu/g^2$ under a large gauge transformation, where
$\mu$ is the coefficient of the CS term and $g$ is the gauge coupling
constant. This leads to the celebrated quantization condition\cite{desjactem}
\beq
q\equiv {4\pi\mu\over g^2}=n,\quad n=0,\pm 1,\pm 2,\cdots
\label{one}
\eeq
in
order to maintain the invariance of $e^{iS}$.

An important question to address is whether this quantization condition is
respected by quantum corrections.  This issue was considered by Pisarski
and Rao\cite{pisrao}, for the case of a pure gauge theory with dynamics
governed by the usual Yang Mills term and the CS term.  They found that
the quantization condition is indeed preserved to one loop; however, the
integer on the  R.H.S.  of (\ref{one}) is shifted by $N$ for ${\cal
G}=SU(N)$.  This calculation has been extended to two loops by Giavirini,
\etal\cite{2loop}, who found no further correction in the limit of pure CS
interaction, confirming the expectation that there are no corrections
beyond one loop in that theory\cite{pisrao}.

Subsequently, the question of quantization was considered for the case of a
completely spontaneously broken gauge theory by Khlebnikov and
Shaposhnikov\cite{khlsha}. They found that at the one loop level $q$ is
multiplicatively renormalized by a complicated function of the three mass
scales in the problem ($\mu$, the symmetry breaking mass scale and the
physical Higgs mass scale), and violates (\ref{one}).

At first sight, this result indicates loss of gauge invariance and
inconsistency of the theory.   However, the situation is not as
catastrophic as it first appears: the underlying theory can continue to
maintain gauge invariance, as manifested for example in the effective
action. The effective action would be a completely gauge invariant
functional of external gauge and scalar fields, but would cease to appear
this way when the scalar field is evaluated at its vacuum expectation
value.   Indeed, it was observed in \cite{khlsha} that in the presence of
spontaneous symmetry breaking, other terms exist  in the effective action
which reduce to the CS term when   $\phi\rightarrow \langle\phi\rangle=v$,
but which are nonetheless {\sl invariant} under large gauge
transformations.  If this is the case, the calculation in \cite{khlsha} is
not, in fact, a calculation of the coeffecient of the CS term alone.
Rather, it is the sum of the coeffecients of the CS term and the other
terms which reduce to it.   This leaves the possibility, which should be
verified, that the  non-quantized result obtained in \cite{khlsha} is the
sum of a quantized CS coeffecient and non-quantized (yet perfectly
acceptable) contributions from the other terms.

In a slightly different context (namely, the spontaneously-broken abelian
case, where quantization of the coefficient of the CS term is not required
for consistency), a complicated radiatively induced correction to
(apparently) the CS term\cite{abssb} was found to be due to other terms in
the effective action which reduce to the CS term in the symmetry-breaking
phase, exactly as described above: the coefficient of the CS term itself is
in fact unchanged in that model\cite{khamacpar}.

In this work, we consider the case of partial breaking of a non-abelian
gauge symmetry. For an abelian unbroken subgroup, this is a proving ground
for the Coleman-Hill theorem\cite{colhil}. The theorem asserts that if
there are no massless states in the theory other than the photon, and if
there is manifest Lorentz invariance, then the
renormalization of the CS term is zero except for the one-loop contribution
of fermions. They
specifically exclude the case of unbroken non-abelian
gauge theories, where each generator forms an abelian subgroup, since there
exists no quantization which respects the conditions of the theorem.
The case of a partially broken non-abelian gauge theory is more interesting.
With
symmetry breaking to just $U(1)$, all the remaining gauge bosons attain
explicit masses; hence, the Coleman-Hill theorem implies no renormalization
of the CS term. On the other hand, for a non-abelian unbroken subgroup
($SU(M)$, where $M<N$,
say), the Coleman-Hill theorem does not apply.
We must still have, at the very least, a quantized coefficient of
the CS term for the gluons of the unbroken subgroup
to have a consistent theory\cite{desjactem}. Although
there is no formal proof, there is a suggestion in \cite{pisrao} that there
might be an equivalent non-abelian version of the Coleman-Hill theorem
which states that under suitable general conditions the renormalization of
the CS term in $SU(N)$ theories is by $q\to q+N$, as found in \cite{pisrao}.
If this is indeed the case, one might
expect that the renormalization of the CS
term for an unbroken, non-abelian sector to be that which corresponds to
the unbroken subgroup only ($q\to q+M$ for unbroken $SU(M)$).
This is because one can construct a Lorentz-invariant gauge where there are
no massless particles outside the unbroken sector, and the conjectured
non-abelian version of the Coleman-Hill theorem (assuming its
assumptions are those of the abelian version) would imply that no
contribution to $q$ will arise from the broken sector.

In the following, we find that exactly this scenario takes place,
to one loop.

We begin by studying the gauge
group $SU(3)$ spontaneously broken to an $SU(2)$ subgroup. The latter being
nonabelian, we must demand the quantization of $\qs$, the value of $q$ for
the unbroken
$SU(2)$ subgroup, including radiative corrections.
(It is important to observe that no terms exist in the effective action
which would reduce under spontaneous symmetry breaking to
the unbroken CS term.) We find,
confirming the work of Chen, \etal\cite{dunne} and correcting the
calculation of the original version of this paper, that
$\qs\to\qrs = \qs+2$.

The coefficient of the CS term remaining quantized in accordance with
(\ref{one}), these
results do not uncover any inconsistency in the quantum version of the
theory of the type discussed in \cite{desjactem}. Nonetheless, we find the
results rather perplexing: the coefficient of the CS term is apparently
a discontinuous function over the phase diagram of the theory.
Specifically, when the pattern of symmetry breaking changes, the
coefficient of the CS term jumps by a discrete amount.
Although such behaviour has been seen before, for example in the case of
massive fermions where the contribution to the CS term is proportional to
the sign of the fermion mass, we do not expect this to arise from scalars.

This behaviour seems to be the rule rather than the exception. Indeed,
we have generalized the above to include
several different patterns of symmetry breaking, and we find in all cases
results consistent with the nonabelian generalization of the Coleman-Hill
theorem conjectured above: when $SU(N)$ is broken to a
subgroup which contains in general $SU(M)$ and $U(1)$ factors, the
renormalization of the CS term for unbroken $U(1)$ subgroups is zero,
while that for unbroken $SU(M)$ subgroups is by $\delta q=M$. The
calculation in all cases separates into contributions from within the
subgroup under consideration and from the orthogonal sector of the theory;
the former gives the simple result just stated while the latter gives a
vanishing contribution.
As for the renormalization of $q$ for the broken sector of the theory, one
can, as outlined above, construct terms in the effective action which
reduce to the CS term when the scalar field is replaced by its expectation
value, and thus the type of calculation undertaken here is insufficient to
determine its renormalization.

We begin with the case of $SU(3)$ spontaneously
broken
to $SU(2)$ via a triplet of Higgs in the fundamental representation.  The
corresponding Lagrangian is given by
\beq
\begin{array}{l}
\LL=-{1\over 4}F_{a\mu\nu }F_a^{\mu\nu }-{\mu\over
2}\epsilon_{\mu\nu\lambda }\left( A_a^\mu\partial^\nu A_a^\lambda
+{1\over 3}gf_{abc}A_a^\mu A_b^\nu A_c^\lambda\right)
\\
\qquad
+\left( D^\mu \phi\right)^\dagger_A\left( D_\mu
\phi\right)_A+m^2\left(\phi^\dagger_A
\phi_A\right)-\lambda\left(\phi^\dagger_A\phi_A\right)^2,
\label{two}
\end{array}
\eeq
where
\beq
\begin{array}{l}
F_{a\mu\nu }=\partial_\mu A_{a\nu}-\partial_\nu A_{a\mu}+gf_{abc}A_{b\mu}
A_{c\nu}
\\
\left( D_\mu \phi\right)_A=\partial_\mu\phi_A-ig{\lambda_{AB}^a\over
2}A_{a\mu}\phi_B
\label{three}
\end{array}
\eeq
and  $\lambda^a_{AB}$ are the Gell-Mann matrices with $f_{abc}$ being the
structure constants.  The choice of the Higgs potential implies a non-zero
vacuum expectation value for $\phi$.  We write
$\phi = \phi^\prime +\langle\phi\rangle_0$
with
\beq
\langle\phi\rangle_0=\pmatrix{0\cr 0\cr {v\over\sqrt2}},
\qquad v=\sqrt{m^2\over\lambda}.
\label{four}
\eeq
The gauge fixing and ghost terms are given by
\beq
\begin{array}{l}
\LL^\prime=-{1\over 2\xi}\left(\partial_\mu A_a^\mu -ig\xi\left(
 \langle\phi\rangle_0^\dagger{\lambda^a\over 2}\phi^\prime
-\phi^{\prime\dagger}
{\lambda^a\over 2}\langle\phi\rangle_0\right)\right)^2
\\
\qquad
+\partial_\mu\bar\eta_a\partial^\mu\eta_a -igf_{abc}\partial_\mu\bar\eta_a
A^\mu_b\eta_c.
\label{five}
\end{array}
\eeq
The vertices are standard for a spontaneously broken non-abelian gauge
theory; however, the CS term introduces an extra (parity odd) three-gluon
vertex.  The gluon propagator is, however, more involved.
It is colour diagonal;
for colour indices in the unbroken sector, $(a=1,2,3)$, in
Landau gauge $(\xi =0)$, it is given by
\beq
i\Delta_{\mu\nu}=-i{(g_{\mu\nu}-{k_\mu k_\nu\over
k^2})-i\mu\epsilon_{\mu\nu\lambda }{k^\lambda\over k^2}\over
k^2-\mu^2},
\label{six}
\eeq
while for the broken generators we have \cite{pisrao,paukha}
\beq
i\Delta_{\mu\nu}=-i{(g_{\mu\nu}-{k_\mu k_\nu\over
k^2})(k^2-m_W^2)-i\mu\epsilon_{\mu\nu\lambda }k^\lambda\over
(k^2-m_+^2)(k^2-m_-^2)}.
\label{seven}
\eeq
Here
\beq
m_\pm=\sqrt{m_W^2+{\mu^2\over4}}\pm{\mu\over 2}
\label{eight}
\eeq
and $m_W$ is the contribution to the gluon mass from the Higgs mechanism.
We have
$m_W=vg/2\equiv m_D$ for the iso-doublet massive vectors
$(a=4,5,6,7)$ and $m_W=vg/\sqrt 3\equiv m_S$ for the iso-singlet massive
vector $(a=8)$.

Following Pisarski and Rao \cite{pisrao}, we calculate $\qr$
according to
\beq
\qr={4\pi\mu\over g^2}Z_m\tilde Z^2
\label{nine}
\eeq
to one loop in Landau gauge, for the unbroken $SU(2)$ subgroup.
Here $Z_m$ and $\tilde Z$ are, respectively, the renormalization constants for
the odd part of the gluon self-energy and the ghost self-energy.
We note that $\qr$ for the broken generators will be different than
that for
the unbroken $SU(2)$ subgroup since the physical Higgs contributes to $Z_m$ in
this case.  We do not present the details of the
calculation since it is amply described in
\cite{dunne}.  For $\qrs$, the renormalized $q$ for
the unbroken $SU(2)$ subgroup,
we must find $Z_m$ and $\tilde Z$ for that sector. Beyond those calculated
in \cite{pisrao}, there are
additional contributions, $\delta Z_m$, to the gluon self-energy coming
from the massive iso-doublet vector bosons circulating in the gluon loop and
from the loop containing unphysical scalars.
The ghost self-energy is also augmented by an additional
contribution $\delta \tilde Z$, from the loop containing massive iso-doublet
vector bosons.  The massive iso-singlet vector boson actually does not
contribute at this order.  We find
\beq
\begin{array}{l}
\delta Z_m=g^2\int{d^3p\over (2\pi )^3}\left(
-2{(p^2+m_D^2)^2\over(p^2+m^2_+)^2(p^2+m^2_-)^2}
+{16\over 3}{p^2(p^2+m_D^2)\over(p^2+m^2_+)^2(p^2+m^2_-)^2}\right.
\\
\qquad
\left.-{2\over 3}{\mu^2(p^2+m_D^2)\over(p^2+m^2_+)^2(p^2+m^2_-)^2}
+2{\mu^2p^2\over(p^2+m^2_+)^2(p^2+m^2_-)^2}
+2{m_D^2\over 3p^2(p^2+m^2_+)(p^2+m^2_-)}\right),
\label{ten}
\end{array}
\eeq
and
\beq
\delta \tilde Z=-{2\over 3}g^2\int{d^3p\over (2\pi )^3}
{(p^2+m_D^2)\over p^2(p^2+m^2_+)(p^2+m^2_-)}.
\label{eleven}
\eeq
To these, we add the Pisarski-Rao contributions (coming from the unbroken
$SU(2)$ sector), yielding:
\beq
Z_m=1+{7g^2\over6\pi m}+\delta Z_m,
\qquad \tilde Z=1-{g^2\over3\pi m}+\delta \tilde Z,
\label{elevena}
\eeq
which, in (\ref{nine}), yields
\beq
\qrs=q+2+q(\delta Z_m+2\delta\tilde Z).
\label{elevenb}
\eeq
The remaining integrals are straightforward; one finds that
$\delta Z_m+2\delta\tilde Z=0$, and
\beq
\qrs=q+2,
\label{elevenc}
\eeq
in agreement with \cite{dunne}.

Some comments are in order. First, $\qrs$ is quantized, as it must
be for consistency of the theory.\footnote{It is perhaps worth reiterating
that computational errors in the original version of this paper led us
to a different conclusion, namely, that $\qrs$ is not quantized.}
Second, there is nonetheless some
peculiar behaviour exhibited.
On the one hand, our final result (\ref{elevenc}) is completely independent
of the expectation value of the scalar field, while on the other hand if we
were to redo the entire calculation in the symmetric phase (in the absence
of spontaneous symmetry breaking), $\qs$ would attain a renormalization
{\sl exactly as in the pure gauge theory with the full gauge group} $SU(3)$:
one would find $\qrs=q+3$.\footnote{Note that, even though the symmetry is
not broken to $SU(2)$ here, we are free to calculate the
radiative correction to $q$ for the gluons of an $SU(2)$ subgroup.} The
limit of symmetry restoration in the above calculation must be examined
with care: we must consider how the limit affects the integrals (\ref{ten})
and (\ref{eleven}) rather than merely studying the final result
(\ref{elevenc}). In fact, the problem can be traced to the third and fifth
terms in the integrand of (\ref{ten}). For instance, the fifth term is
simply not present in the symmetric phase since there is no
gluon-gluon-scalar vertex (the vertex is proportional to $v$, whose
presence in the fifth term is contained in $m_D$), while the symmetric
limit ($v\to0$)
of the integral of this term is nonzero. This ambiguity is due to an
infrared problem which appears in the integrand as $v\to0$. In this limit,
$m_-\to0$, and the integral is linearly divergent. That term's
contribution to $\delta Z_m$ is of the order $g^2 {m_D}^2/m_-$. Since
$m_D\propto v$ while as $v\to0$ $m_-\sim v^2$ (as can be seen from
(\ref{eight})), the contribution of that term is finite and nonzero
as $v\to0$, in disagreement with the zero result one would have obtained
obtained in the symmetric theory. Similar considerations apply to the third
term, while the other terms in (\ref{ten}) and (\ref{eleven}) are
well-behaved in the symmetric limit. Thus, we conclude that
evaluating the integral and
taking the limit of no symmetry breaking do not commute. Non-commutativity
of limits has also been observed in perturbative calculations in CS
theories in several other situations \cite{khlsha,abssb,lebtho}.

The calculations outlined above can be easily modified to handle other
cases. We have studied the following patterns of symmetry breaking:
$SU(2)\to U(1)$ via a real triplet which attains an expectation value
$\langle\phi_a\rangle=v\delta_{a,3}$;
$SU(3)\to SU(2)\times U(1)$ via an adjoint which attains an expectation
value $\langle\Phi\rangle=v T_8\sim diag(1,1,-2)$;
$SU(3)\to U(1)\times U(1)$ via an adjoint which attains an expectation
value $\langle\Phi\rangle=v T_3\sim diag(1,-1,0)$.
In all cases, the Feynman rules are found in a straightforward way, and the
diagrams which contribute to $Z_m$ and to $\tilde Z$ for an unbroken gluon
are identical to those as calculated above. Differences arise only in the
values of coupling constants, masses, and group theoretical factors.
The calculation of the $Z$s naturally separates into contributions from
the unbroken group and possible contributions from the orthogonal (broken)
sector. In the case of an unbroken group which is a direct product, $\delta
q$ can be computed for each subgroup of the direct product. Straightforward
group theoretical factors imply that the subgroups decouple: each subgroup of
the unbroken group only contributes to its own $\delta q$. As for the
contribution from the broken sector, in all
cases it was found that
$\delta Z_m+2\delta\tilde Z=0$, and the net result was that $\delta q$ is
that value one would have calculated from the pure gauge theory of the
unbroken subgroup under consideration. Thus, in all cases the correction to
$\qu$ was zero, in keeping with expectations based on the
Coleman-Hill theorem \cite{colhil}. Furthermore, in
the second case, the unbroken $SU(2)$ correction is as above: $\delta
\qs=2$. The generalization is clear: the $q$ of any
residual $U(1)$ symmetry receives no radiative correction,
while that of any residual non-abelian group receives a radiative
correction which is as if the orthogonal sector of the theory
was not there.

\bigskip\bigskip
We thank A.S. Goldhaber,
M. Leblanc, G. W. Semenoff and V. P. Spiridonov for useful
discussions. We are particularly indebted to G. Dunne for discussions of
his results which enabled us to locate a couple of errors in the
original version of this work.
A. Khare thanks the Laboratoire de Physique Nucl\'eaire for the
kind invitation and hospitality during his visit.  This work supported in part
by NSERC of Canada and FCAR du Qu\'ebec.


\begin{references}

\bibitem{1}
W. Siegel,\np B156, 135, 1979;
J. Schonfeld, \np B185, 157, 1981;
R. Jackiw and S. Templeton, \prd 23, 2291, 1981.

\bibitem{desjactem}
S. Deser, R. Jackiw and S. Templeton, \annp 140, 372, 1982.

\bibitem{any}
J. M. Leinaas and J. Myrheim, \journal Nuovo Cimento, 37B, 1, 1977;
F. Wilczek,\prl 48, 1144, 1982.

\bibitem{csterm}
D.P. Arovas, \etal, \np B251 [FS13], 117, 1985;
F. Wilczek and A. Zee, \prl 51, 2250, 1983.

\bibitem{fqhe}
R. Laughlin, \prl 50, 1395, 1983;
B.I. Halperin, \prl 52, 1583, 1984;
S. C. Zhang, T. H. Hansson and S. Kivelson, \prl 62, 82, 1989, and references
therein.

\bibitem{anysup}
R. Laughlin, \journal Science, 242, 525, 1988;
Y. H. Chen, B.I. Halperin, F. Wilczek and E. Witten, \ijmp B3, 1001, 1989.

\bibitem{purecs}
C.R.
Hagen, \annp 157, 342, 1984.

\bibitem{topol}
E. Witten, \cmp 121, 351, 1989.

\bibitem{2dg}
E. Witten, \np B311, 46, 1988.

\bibitem{hit}
S. Weinberg, in ``Understanding the Fundamental
Constituents of Matter", editor A. Zichichi (Plenum Press, NY, 1978);
A. Linde, \journal Rep. Prog. Phys., 42, 389, 1979.

\bibitem{pisrao}
R. D. Pisarski and S. Rao, \prd 32, 2081, 1985.

\bibitem{2loop}
G. Giavirini, C. P. Martin and F. Ruiz Ruiz, Nucl. Phys. {\bf B381},
222, 1992.

\bibitem{khlsha}
S. Yu. Khlebnikov and M. E. Shaposhnikov, \pl 254B, 148, 1991.

\bibitem{abssb}
S. Yu. Khlebnikov, \journal JETP Letters, 51, 81, 1990
V. P. Spiridonov, \journal JETP Letters, 52, 1112, 1990.

\bibitem{khamacpar}
A. Khare, R. MacKenzie and M.B. Paranjape, \pl B343, 239-243, 1995.

\bibitem{colhil}
S. Coleman and B. Hill, \pl 159B, 184, 1985; see also the discussion in
\cite{pisrao}.

\bibitem{dunne}
L. Chen, G. Dunne, K. Haller and E. Lim-Lombridas,
\pl 348B, 468, 1995.

\bibitem{paukha}
S. K. Paul and A. Khare,\pl 171B, 244, 1986.

\bibitem{lebtho}
M. Leblanc and M. T. Thomaz, \pl 281B, 259, 1992.

\end{references}
\end{document}